\documentclass{PTI-KKIO}

\usepackage{url}
\usepackage[english]{babel}
\usepackage[utf8]{inputenc}
\usepackage{graphicx}
\usepackage[T1]{fontenc}
\usepackage{xr}
\usepackage{listings}
\begin{document}

\begin{frontmatter}

\setcounter{page}{1}
\booktitle{\scriptsize{Agnieszka Patalas, Wojciech Cichowski, Michał Malinka, Wojciech Stępniak, Piotr Maćkowiak, and Lech Madeyski, “Software Metrics in Boa Large-Scale Software Mining Infrastructure: Challenges and Solutions” in Software Engineering: Improving Practice through Research (B. Hnatkowska and M. Śmiałek, eds.), pp. 131–146, 2016.}}
\bookeditors{\tiny{URL: http://madeyski.e-informatyka.pl/download/Patalas16.pdf}}

\title{Software Metrics in Boa Large-Scale Software Mining Infrastructure: Challenges and Solutions}

\author[A]{\fnms{Agnieszka} \snm{Patalas}},
\author[A]{\fnms{Wojciech} \snm{Cichowski}},
\author[A]{\fnms{Michał} \snm{Malinka}},
\author[A]{\fnms{Wojciech} \snm{Stępniak}},
\author[A]{\fnms{Piotr} \snm{Maćkowiak}}
and
\author[A]{\fnms{Lech} \snm{Madeyski}}

\runningauthor{Patalas et al.}
\runningtitle{Software Metrics in Boa Large-Scale Software Mining Infrastructure}

\address[A]{Wrocław University of Science and Technology, Faculty of Computer Science and Management, Wrocław, Poland}

\begin{abstract}
In this paper, we describe our experience implementing some of classic software engineering metrics using Boa---a large-scale software repository mining platform---and its dedicated language. We also aim to take an advantage of the Boa infrastructure to propose new software metrics and to characterize open source projects by software metrics to provide reference values of software metrics based on large number of open source projects. Presented software metrics, well known and proposed in this paper, can be used to build large-scale software defect prediction models. Additionally, we present the obstacles we met while developing metrics, and our analysis can be used to improve Boa in its future releases. The implemented metrics can also be used as a foundation for more complex explorations of open source projects and serve as a guide how to implement software metrics using Boa as the source code of the metrics is freely available to support reproducible research.
\end{abstract}

\end{frontmatter}

\thispagestyle{copyright}
\pagestyle{headings}

\section{Introduction}
Boa is a tool that can be used for data mining repositories of open-source projects. It contains the full history of a repository---from every revision's date and author, data on added, deleted and modified files to the complete state of the repository at the moment of commit. All data can be obtained by using the dedicated language. Boa provides a set of functions, which can be used for advanced data filtering \cite{boaDocumentation}\cite{dyer2013}.

Boa has already been used for a variety of studies, including developers' willingness to adapt new Java features \cite{dyer2014} or the licenses used in open-source projects \cite{vendome2015license}. So far they have not been metrics-oriented, even though the tool is intended to be used this way, as implicated by the inclusion of appropriate examples in the documentation of Boa \cite{boaExamples} (e.g., {\em What are the number of attributes (NOA), per-project and per-type?}, {\em What are the number of public methods (NPM), per-project and per-type?}).

In this paper we focus on using Boa infrastructure to answer three research questions:
\begin{enumerate}
\item Which of the classic, widely known, software engineering metrics can be implemented in Boa?

The implementation of classic software engineering metrics in Boa and publication of calculation scripts will make it easier to extend existing small-scale empirical software engineering research using software metrics, performed usually on a small number of projects, to a large-scale research.

\item What new metrics, that take advantage of the Boa's unique infrastructure, can be proposed?

This paper will serve as a guide, for other researchers and practitioners, which shows how to implement new software metrics taking into account the unique features, as well as limitations, of the Boa large-scale software repository mining platform.

\item What is the feasibility of defect prediction models based on large number of projects data obtained from Boa data sets?

According to our knowledge, this is one of the first attempts (if not the first) to build large-scale software defect prediction models based on a very large number of projects. Existing software defect prediction models usually base on a very limited number of projects.

\end{enumerate}
Presented study refers to state of Boa framework during October 2015 - January 2016 period -- when the source material was gathered.

\section{Research methodology} \label{sec:Methodology}
In this section we introduce briefly into the following topics: how we selected projects for further investigation (see Section~\ref{sec:ProjectsSelection}), how we implemented software metric scripts using the Boa language (see Section~\ref{sec:ImplementationOfSEMetrics}), and how we built software defect prediction models using software metrics from Boa (see Section~\ref{sec:DefectPredictionModel}), including also how we obtained data from the Boa output files (see Section~\ref{sec:weka}). 

\subsection{Projects selection}\label{sec:ProjectsSelection}
Boa source code described in this paper has been developed and tested on two Boa data sets: September 2015 GitHub, and September 2013 SourceForge. A special filtering has been applied to select projects passing some entry criteria.
The software projects explored in our study had to pass the following criteria:
\begin{enumerate}

\item \textbf{They have to have a code repository with revisions.} The \textit{2013 September/SourceForge} data set consists of 700k projects. Our analysis with Boa queries has shown that 30\% of them have no code repository ~\cite[Section 2.1]{onlineAppendix}. Out of remaining 489k (amount close to this stated by Boa developers - 494,158 \cite{boaStatistics}) 4,767 projects have two repositories. Repositories in those projects have common history of revisions~\cite[Section 1.1]{onlineAppendix}. In case of projects with multiple repositories, only the first of them is considered during study to avoid data duplication. Out of 489k projects with one code repository, 423k of them had no code revisions (commits)~\cite[Section 1.2]{onlineAppendix}. It is difficult to determine whatever or not Boa is missing some data---the data sets have been defined for a given month in a given year, and current state of the repository might be different.

The \textit{2015 September/GitHub} data set has 7.83 million projects. 95\% of them have no code repository in the Boa framework, even though the majority of them is available from the GitHub website. They are active and public, but most of them have had no commits since 2013 \cite[Section 1.3]{onlineAppendix}. From 380k projects with repository, only 2486 of them had commits in 2015 \cite[Section 1.4]{onlineAppendix}. Out of the entire GitHub dataset, 4\% of projects have code repositories with revisions~\cite[Section 1.5]{onlineAppendix}.

\item \textbf{They have to have over 100 commits.} The projects picked should be mature enough for metrics calculation. A larger number of commits usually means a larger number of \textit{fixing revisions}, which are in turn used for development of software defect prediction models.

\item \textbf{They have to be written in Java.} Java has been picked for this research due to being a mature, object-oriented language, popular among developers. 
It is also worth mentioning that Boa is written in Java, as well as provides extra Java-specific options, such as recognizing Java source files with and without parsing errors.
\end{enumerate}

The Boa language implementation of filters to select projects fulfilling the above mentioned criteria is presented in Listing \ref{lst:filtersBoaImpl}.

\begin{lstlisting}[captionpos=b, caption=Implementation of filters, label={lst:filtersBoaImpl}, basicstyle=\footnotesize]
before node: Project -> {
  # They have to be written in Java.
  ifall (i: int; !match(`^java$`,
    lowercase(node.programming_languages[i]))) stop; 

  # They have to have a code repository with revisions.
  if(len(node.code_repositories) > 0) {
    visit(node.code_repositories[0]);
  }
  stop;
}
before node: CodeRepository -> {
  # They have to have over 100 commits.
  if(len(node.revisions) < 100) stop;
  ...
}
\end{lstlisting}

The final number of projects that passed our entry criteria is presented in Table~\ref{table:datasets}.

\begin{table}[ht!]
\centering
\caption{Data sets}
\begin{tabular}{|l|l|l|}
\hline
\textbf{Dataset} & \textbf{All projects} & \textbf{Accepted projects}\\ \hline
GH small & 7,988 & 29\\ \hline
GH medium & 783,982 & 2485\\ \hline
GH large & 7,830,023 & 25307\\ \hline
SF small & 7,029 & 50\\ \hline
SF medium & 69,735 & 666\\ \hline
SF large & 699,331 & 7407\\ \hline
\end{tabular}
\label{table:datasets}
\end{table}

\subsection{Implementation of SE metrics}\label{sec:ImplementationOfSEMetrics}
All of the metrics are calculated for classes.
Each of the metric is implemented as a different Boa query, and is run on all Boa data sets mentioned in Section~\ref{sec:Methodology}. Due to long execution time, only data from GH small and SF small data sets are used for creating prediction models later on.

The output file of a query has to have the following data:
\begin{itemize}
\item{the ID of the project}
\item{the ID of the class}
\item{the value of the calculated metric or the expected value}
\end{itemize}
This approach makes it possible to effortlessly merge all values gathered as the outputs of Boa queries, so they can be used as an input data set for a prediction model.

\subsection{Defect prediction model}\label{sec:DefectPredictionModel}
Software defect prediction model is aiming to find the classes that cause the most defects. A simple strategy to find them is searching for the classes that had been fixed most frequently.
\subsubsection{Expected value - NCFIX} \label{sssec:expValue}
The expected value in our defect prediction model is Number of Class Fixes.
Based on Boa’s abilities, it is assumed the class has been fixed, if the two following conditions have been met:
\begin{itemize}
\item  the file containing the class has been modified in a revision;
\item  the revision is marked as a fixing revision by the Boa’s function\\ \texttt{isfixingrevision} \cite{boaDocumentation}.
\end{itemize}
The list of classes and their fixes is obtained by the following algorithm:

\begin{enumerate}
\item Create an empty key-value collection for storing respectively: files in projects, number of fixing revisions for each file.
\item Visit a project's repository revision.
\item Check if it’s a fixing revision.
\item Investigate the files changed in this revision.
\begin{enumerate}
\item If a file is marked as deleted, remove it from the collection.
\item If a file is added to the project in the current revision, add it to the collection:
\begin{enumerate}
\item with a value of 1 if the revision is a fixing one;
\item with a value of 0 otherwise.
\end{enumerate}
\item If a file is modified in the current revision, update it in the collection
\begin{enumerate}
\item increment the number of fixes by one, if the revision is a fixing one;
\item leave it otherwise.
\end{enumerate}
\end{enumerate}
\item Repeat steps 2-4 until you reach the most recent revision and there is no more revisions to check.
\item For all files stored in the collection, select only the ones that declare classes. Return the identifiers of the classes, and numbers of fixes corresponding to their files as the output.
\end{enumerate}
The algorithm is inspired by the \texttt{getsnapshot} function implemented by Boa \cite{boaDocumentation}, which returns the state of the repository at given time stamp.

\subsection{The use of Boa API and Weka}\label{sec:weka}
To allow easy management of Boa jobs and connecting job outputs with development of defect prediction models, a simple Java program \cite{sourceCodeGithub} has been written. The software uses Boa Java API \cite{boaAPI} release 0.1.0 to run jobs. Data from Boa is transformed into \texttt{.arff} file of following format:
\begin{lstlisting}[basicstyle=\footnotesize]
@RELATION classes
@ATTRIBUTE classID  string
@ATTRIBUTE M_1   NUMERIC
...
@ATTRIBUTE M_N   NUMERIC
@ATTRIBUTE fixingRevisions   NUMERIC
\end{lstlisting}
where \texttt{classID} is an identifier of a studied class; \texttt{M\_1 ... M\_N} is a vector of calculated metrics for a class from latest repository SNAPSHOT; \texttt{fixingRevisions} attribute is the expected value described in Section~\ref{sssec:expValue}.

\section{Results}
In this section three kinds of contribution are discussed, related to implementation of classic and new software metrics in Boa, as well as development of software defect prediction models on a basis of very large number of software projects. The latter can be seen as a way to address external validity threats common for most of the the empirical studies focused on software defect prediction.
All metrics' implementations are available to download via links provided in appendix \cite[Section 3]{onlineAppendix}.

\subsection{Implementation of classic software engineering metrics}
This section presents how to implement scripts to collect some of the well-known, classic software metrics~\cite{chidamber1994metrics} in Boa. The metrics were chosen based on their popularity and Boa's limitations.

\subsubsection{Obtaining classes}

Using \texttt{getsnapshot} function implemented in Boa, all files available in the most recent revision of the project are gathered. Then, they are filtered so that only the files containing classes are taken into consideration. The data stored in the \texttt{Declaration}~\cite{boaDocumentation} and its attributes are used for calculating the value of a metric.

\subsubsection{Inheritance issue}

Each declaration (class or interface) node in Boa has its array of parents \cite{boaDocumentation}. However, those parents are presented only as \textit{Types}, meaning, they only have \textit{TypeKind} (determining if it’s a class, interface, or something else) and name, without its full package path or any other identifier.
If two classes or interfaces in a project have the same name, but they are in different packages, it is impossible to determine which one is the ancestor of a given declaration.
Therefore, all metrics using inheritance (such as all of the MOOD metrics \cite{moodMetrics}, Depth of Inheritance Tree, Number of Children and Coupling between Object Classes~\cite{chidamber1994metrics}) had to be, unfortunately, excluded from the study.

\subsubsection{Metrics obtained directly from the \texttt{Declaration} node}

Weighted Methods per Class (WMC) in its base version---the sum of methods in a class, Number of Fields (NoF) and Number of Nested Declarations (NoND), presented in Table~\ref{table:declarationmetrics}, have been successfully implemented using the structure of the \texttt{Declaration} node alone.

\begin{table}[ht!]
\centering
\caption{\texttt{Declaration} attributes and associated metrics}
\begin{tabular}{l|c}
\textbf{Attribute} & \textbf{Metric}\\ \hline
methods & WMC\\ \hline
fields & NoF\\ \hline
nested\_declarations & NoND\\
\end{tabular}
\label{table:declarationmetrics}
\end{table}

For each of those metrics, the value is a length of the attribute array.
The execution time for those metrics is relatively small, up to 10 minutes for the biggest data sets, which clearly shows the advantages of using Boa and the approach to calculate metrics using the structure of the \texttt{Declaration} node, presented in this paper.

\subsubsection{Response For a Class (RFC)}
The RFC metric was implemented as a number of methods in the class, added to number of remote methods directly called by methods of the class.
\newline
The issue with the implementation of this metric is that Boa makes it difficult to recognize the difference between class’ inner method and method of the external classes of the same identifier.
For example: the method \texttt{getId()} of \texttt{class A}, called in \texttt{class B}, is seen as the same as method \texttt{getId()} in \texttt{class B}. If \texttt{class A} called two methods of the same name from different classes (\texttt{class B} and \texttt{class C}), those would be indistinguishable as well.
\newline
There is no direct method that would allow to instantly determine the types of called methods' arguments \cite{boaDocumentation} as well as the type of instance of variable from which the method was called \cite[Section 1.6]{onlineAppendix}. Such information can be obtained only by deeper analysis of Boa's AST tree, to the level of single \texttt{Statement}s.

The simplified version of the metric, that ignores this nuance, has been successfully implemented and ran for both Boa's data sets.

\subsection{Implementation of new software metrics}
The metrics presented below have been developed by us upon learning more about the Boa architecture and its tree structure.

\subsubsection{Number of Statements in Methods}

The NoSiM metric is calculated as a sum of all statements in class methods. The nodes calculated are of the Boa type \texttt{Statement}. For studied Java classes, those nodes are either blocks of code marked by \texttt{\{\}} or single code expressions.
The implementation of this metric is a starting point for implementation of a Lines of Code (LoC) metric. To achieve the LoC metric, all class' fields, number of methods, and such, would have to be added.

\subsubsection{Maximum Depth of Declaration Nesting} \label{sssec:MDoDN}
MDoDN is the maximum level of class nesting in a class.
For the following code:
\begin{lstlisting}[basicstyle=\footnotesize]
class A {
    class B {
        class C {}
    }
    class D {}
}
\end{lstlisting}
the result for \texttt{class A} would be 3 (the depth of \texttt{C} class). The metric is not calculated for nested classes (in the example: \texttt{B, C,} and \texttt{D}). For implementation of this metric, Boa's stack functions are used. Every time the node of a nested \texttt{Declaration} is entered, it is pushed onto the stack. The metric value is the stack's element count.

\subsubsection{Number of Anonymous Declarations}
NoAD for Java is a sum of all anonymous children classes in the parent class.
To calculate this metric, the \texttt{Expression} Boa node is tested for having a \texttt{Declaration} with a parameter of \texttt{ANONYMOUS} type.

\subsubsection{Cumulative metrics}
\label{sssec:CumulativeMetrics}
Metrics NoM, NoF, NoSiM, NoAD and NoND have been also successfully implemented in cumulative versions (CNoM \cite[Section~1.7]{onlineAppendix}, CNoF \cite[Section~1.8]{onlineAppendix}, CNoSiM \cite[Section~1.9]{onlineAppendix}, CNoAD \cite[Section~1.10]{onlineAppendix}, CNoND \cite[Section~1.11]{onlineAppendix}), where calculated value is a sum of metric for not only a class, but also all its nested and local classes.

\subsection{Defect prediction model}\label{subsec:predictionModel}
The defect prediction model presented below is a single defect prediction model calculated for a high number of Boa projects. This is different from a traditional approach, with a single, or several projects used to develop defect prediction models.

Data obtained from the Boa output files (described in Section~\ref{sec:weka}) is randomly separated into training set and testing set (in 9:1 proportion). The \texttt{fixingRevisions} attribute in the testing set is nulled out, so it can be calculated using prediction model.

We used Random Forest to build defect prediction model. Random Forest generates a lot of random samples which are the subsets of training data set. A decision tree is generated for each of the samples \cite{Breiman2001}. The parameters listed below have been determined experimentally:
\begin{itemize}
	\item {number of trees : 200},
	\item {max depth : 12},
	\item {number of features : 12},
	\item {cross-validation folds: 10},
	\item {random seed: 1}
\end{itemize}
The results of 10-fold cross-validation are presented in Table~\ref{table:evalModelResults}. Pearson product-moment correlation coefficient $r$ shows a low correlation between the results from defect prediction model and real values, with high error ratio. Those results are further analyzed in Section \ref{sec:dicussion}. 

\newcommand{\specialcell}[2][c]{\begin{tabular}[#1]{@{}c@{}}#2\end{tabular}}
\begin{table}[ht!]
\centering
\caption{Results of evaluation of the prediction model}
\begin{tabular}{l|c c}
\textbf{Evaluation attribute} & \textbf{\specialcell{GH 2015\\(small)}} & \textbf{\specialcell{SF 2013\\(small)}}\\ \hline
Correlation coefficient (R) & 0.215 & 0.244\\ \hline
Mean absolute error (MAE) & 2.16 & 0.603\\ \hline
Root mean squared error (RMSE) & 9.96 & 1.32\\ \hline
Relative absolute error (RAE) & 102\% & 93.3\%\\ \hline
Root relative squared error (RRSE) & 100\% & 97.8\%\\ \hline
\end{tabular}
\label{table:evalModelResults}
\end{table}

\subsection{Reference values of software metrics}

The subsequent goal was to characterize a large number of open source projects available from Boa by means of software metrics in order to create reference values of software metrics.
Table~\ref{table:stats} presents descriptive statistics for each of calculated metrics among the data sets.

\begin{table}[ht!]
\centering
\caption{Mean, median and standard deviation for metrics calculated in the study.}
\resizebox{\textwidth}{!}{
\begin{tabular}{l|c|c|c|c|c|c|c|c|c|c|c|c}
\textbf{\specialcell{Metric}} & \multicolumn{3}{|c|}{\textbf{\specialcell{GH2015\\(all)}}} & \multicolumn{3}{|c|}{\textbf{\specialcell{GH2015\\(fixes > 0)}}} & \multicolumn{3}{|c|}{\textbf{\specialcell{SF2013\\(all)}}} & \multicolumn{3}{|c|}{\textbf{\specialcell{SF2013\\(fixes > 0)}}}\\ \hline
\textbf{\specialcell{ }} & \textbf{\specialcell{$\mu$}} & \textbf{\specialcell{$\tilde{x}$}} & \textbf{\specialcell{$\sigma$}}  & \textbf{\specialcell{$\mu$}}  & \textbf{\specialcell{$\tilde{x}$}} & \textbf{\specialcell{$\sigma$}} & \textbf{\specialcell{$\mu$}} & \textbf{\specialcell{$\tilde{x}$}} & \textbf{\specialcell{$\sigma$}}  & \textbf{\specialcell{$\mu$}}  & \textbf{\specialcell{$\tilde{x}$}} & \textbf{\specialcell{$\sigma$}} \\ \hline
NOAD & 0.12 & 0 & 0.86 
& 0.15 & 0 & 1.05 
& 0.17 & 0 & 1.31
& 0.34 & 0 & 2.40\\ \hline
CNOAD & 0.13 & 0 & 0.93 
& 0.17 & 0 & 1.13
& 0.19 & 0 & 1.39
& 0.36 & 0 & 2.51\\ \hline
NOND & 0.32 &  0 & 1.20 
& 0.32 & 0 & 1.28
& 0.14 & 0 & 0.83
& 0.22 & 0 & 1.00\\ \hline
NOF & 2.98 & 1 & 15.64  
& 3.10 & 1 & 9.29
& 3.75 & 2 & 8.51
& 4.36 & 2 & 11.43\\ \hline
CNOM & 7.10 & 3 & 18.69 
& 7.50 & 4 & 13.84
& 8.97 & 5 & 14.21
& 11.48 & 6 & 19.27\\ \hline
MDODN & 0.20 & 0 & 0.44 
& 0.22 & 0 & 0.47
& 0.14 & 0 & 0.38
& 0.20 & 0 & 0.45\\ \hline
CNOSIM & 40.33 & 13 & 104.15
& 48.51 & 16 & 126.33
& 65.84 & 26 & 177.32
& 90.83 & 33 & 221.83\\ \hline
NOSIM & 36.80 & 13 & 92.57
& 44.43 & 15 & 116.30
& 61.58 & 24 & 170.26
& 83.28 & 31 & 208.54\\ \hline
NOM & 6.40 & 3 & 16.85
& 6.67 & 3 & 11.64 
& 8.15 & 5 & 12.41
& 10.12 & 5 & 17.09\\ \hline
CNOF & 7.10 & 3 & 18.69
& 7.50 & 4 & 13.84
& 8.97 & 5 & 14.21 
& 11.48 & 6 & 19.27 \\ \hline
CNOND & 0.33 & 0 & 1.26
& 0.34 & 0 & 1.37
& 0.14 & 0 & 0.87
& 0.23 & 0 & 1.05\\ \hline
RFC & 13.03 & 7 & 23.27 
& 15.19 & 8 & 22.55
& 19.06 & 12 & 24.10
& 24.81 & 15 & 32.75\\
\end{tabular}
}
\label{table:stats}
\end{table}

\section{Discussion}
\label{sec:dicussion}

The presented prediction model was tested on small data sets, but with correct resources it can be easily scaled to use full data sets with up to 25k subjects. This use case would be, to the best of our knowledge, the first attempt to create a large-scale defect prediction model, as other examples from literature show prediction models developed using less than 200 projects \cite{JureczkoMadeyski10,JureczkoMadeyski15,Madeyski15SQJ}.

The performance of the prediction model is poor due to the fact that a majority of classes studied has zero fixing revisions and therefore input data is highly unbalanced, see Table~\ref{table:classFixes}.
However, the quality of prediction model and employing methods to deal with the class imbalance problem are not the main objectives of the study.
Our aim was to show that it is possible to collect all the data necessary to build a large-scale software defect prediction model using the Boa platform.

Results from Table~\ref{table:stats} show that not for all metrics standard deviation is lower for filtered datasets. This can be caused by the nature of metrics (such as NoND, NoAD, MDoDN), which are unlikely to have a high mean value in majority of projects.

\begin{table}[ht!]
\centering
\caption{Number of classes with zero and more than zero fixes in datasets}
\begin{tabular}{c|c|c}
\textbf{\specialcell{Amount of\\class fixes}} & \textbf{\specialcell{GH 2015\\small}} & \textbf{\specialcell{SF 2013\\small}}\\ \hline
0 & 13296 (58.9\%) & 30244 (80.1\%)\\ \hline
\textgreater 0 & 9260 (41.1\%) & 7504 (19.9\%)\\ 
\end{tabular}
\label{table:classFixes}
\end{table}

\subsection{Further research}
It is worth to look at the way the fix in the revision is identified. Boa-provided function \texttt{isfixingrevision} is based only on the commit message text analysis. We assume this function is not ideal and integrating Boa API with outside software, such as bug tracking systems, can be a better solution to determine existing bugs in code revisions.

The data used for building prediction models in our study has big disproportions. Applying different filters and criteria (more mature projects, different languages and so on) could provide better data set for analysis, with more fixing revisions.

An interesting path of further research are process metrics~\cite{Madeyski15SQJ,Madeyski11}, which reflect changes over time and are becoming the crucial ingredients of software defect prediction models.

\section{Conclusions}
Overall, the goal of the research, as described with research questions --- implementation of software metrics in Boa and collecting data sets from a large number of projects, e.g., for the sake of prediction models ---has been achieved. We were able to implement some of the classic software engineering metrics using Boa, we presented some Boa-specific metrics, and we made an attempt to create a defect prediction model with the data we gathered. This proves that Boa can be a useful tool for data mining analysis in this particular field, as well as for creating sophisticated queries regarding its data sets. However, Boa is still a new framework that comes with a few disadvantages, and some of the metrics and operations were impossible to implement at the moment. In the following sections, the challenges met and our solutions are presented.

\subsection{Challenges}
\label{subsection:challenges}
Boa uses visitor pattern---one of Boa's greatest strengths---which sometimes might provide unexpected results if queries are not written properly.

\subsubsection{Local and nested classes}
One of the first issues we encountered creating Boa queries was a different size of output jobs. For our metrics, we gathered all classes from all projects. Therefore, for the same data set, all queries should return the same number of rows.
As it turned out, the difference was caused by the behaviour of the visitor pattern, used by Boa. When source code contains a local class (class defined inside one of the methods) or a nested class (a class declared inside of another class), this class is visited by the visitor pattern before the analysis of the class containing it ends. Upon returning to the class-container, some of it's metrics and calculations had been assigned to the local or nested class.

\textbf{Solution:} Boa offers implementation of stacks, which we started using while visiting local and nested classes. We took advantage of this solution implementing the Maximum Depth of Declaration Nesting metric described in Section~\ref{sssec:MDoDN}.

\subsubsection{Boa code compilers}
Boa uses two different code compilers for SourceForge and GitHub data sets. As the framework is still in early development, sometimes the same query acts differently depending on the data set used.

\textbf{Example:} One of Boa sample queries "How many committers are there for each project?" \cite{boaExampleCommiters} works fine in SF \cite[Section 1.12]{onlineAppendix}, but causes compilation error in GH \cite[Section 1.13]{onlineAppendix}.
\newline
In that case, a small change in the code notation solved the issue \cite[Section 1.14]{onlineAppendix}:

\begin{itemize}
    \item Code resulting with error:
\begin{lstlisting}[basicstyle=\footnotesize]
committers[p.code_repositories[i].revisions[j].committer.username] = true;
\end{lstlisting}
    \item Code resulting with success:
\begin{lstlisting}[basicstyle=\footnotesize]
username : string = p.code_repositories[i].revisions[j].committer.username;
committers[username] = true;
\end{lstlisting}
\end{itemize}

This example shows that a person creating queries with Boa might run into different issues depending on the data set picked.

During our research, we often used Boa dictionaries. Dictionaries are defined by Boa as \texttt{map[key\_type] of [value\_type]}. Boa returns an error, if \texttt{int} is used as a \texttt{value\_type}. We must have stored our integer values as strings, which resulted in converting value to integer each time it was used in calculations, and then back to string to update the map.

\subsubsection{Debugging process}
The errors reported by Boa are often lacking any sort of description. The debugging process comes down to commenting out parts of queries to check which fragments are causing errors. Each code test takes about a minute (and then some follow-up time to check if the output data is correct), and sometimes multiple tests are required to find the source of an error. There is no way of tracking the execution of the queries.

\textbf{Solution:} All variables used during the debugging process have to be initiated, by defining its type and aggregation method, and then returned in the output file.

\subsection{Contribution}
The paper describes our experience with using Boa platform for implementing software engineering metrics and defect prediction models. Our findings can be useful for both researchers---with solutions presented in Section~\ref{subsection:challenges} and provided source codes for metrics we implemented---as well as developer teams and project managers, providing an example for obtaining large-scale SE metrics for projects of particular profile (i.e. number of commits, used programming language and so on).
The metric implementations proposed by us are scalable---calculated for classes, but could be as well implemented for packages or projects.

Based on our findings, we confirm that Boa can be a powerful data mining tool, which can be used for a variety of research, alone and with usage of other software, like Weka, as demonstrated in Section~\ref{sec:weka}.

\end{document}